\documentclass[aps,prb,twocolumn,showpacs,showkeys,preprintnumbers,groupedaddress,amsmath,amssymb,]{revtex4-1}

\usepackage{graphicx}
\usepackage{dcolumn}
\usepackage{bm}
\usepackage{hyperref}
\usepackage[dvips]{color}

\begin{document}

\title{Logarithmic temperature dependence of Hall transport in granular metals}

\author{Yu-Jie Zhang$^1$}
\author{Zhi-Qing Li$^1$}
\email[Electronic address: ]{zhiqingli@tju.edu.cn}
\author{Juhn-Jong Lin$^{2,}$}
\email[Electronic address: ]{jjlin@mail.nctu.edu.tw}

\affiliation{$^1$Tianjin Key Laboratory of Low Dimensional Materials Physics and
Preparing Technology, Faculty of Science, Tianjin University, Tianjin 300072,
China \\ $^2$Institute of Physics and Department of Electrophysics, National Chiao Tung University, Hsinchu 30010,
Taiwan}

\date{\today}

\begin{abstract}

We have measured the Hall coefficient $R_H$ and the electrical conductivity
$\sigma$ of a series of ultrathin indium tin oxide films between 2 and 300 K.
A robust $R_H$\,$\propto$\,ln$T$ law is observed in a considerably wide
temperature range of 2 and $\sim$120 K. This ln$T$ dependence is explained as
originated from the electron-electron interaction effect in the presence of
granularity, as newly theoretically predicted. Furthermore, we observed a
$\sigma$\,$\propto$\,ln$T$ law from 3 K up to several tens K, which also arose
from the Coulomb interaction effect in inhomogeneous systems. These results
provide strong experimental supports for the current theoretical concepts for
charge transport in granular metals with intergrain tunneling conductivity
$g_T$$\gg$1.

\end{abstract}

\maketitle

Granular metals are composite materials in which the metallic
granules are randomly embedded in an insulating matrix. Recently, the
electronic conduction properties of granular metals have attracted much
renewed theoretical \cite{zhang1,zhang2,zhang3,zhang4}
and experimental \cite{zhang5,zhang6,Huth09,zhang10,Gondorf11} attention, due to the
improved nanoscale feature and rich fundamental phenomena in the presence of
structural inhomogeneities. In particular, the \textit{intragrain} electron dynamics
is found to play a crucial role in the Hall transport \cite{zhang4} which has
often been overlooked in the everlasting studies of granular systems.
Previously, great efforts have long been focused on the \textit{intergrain} electron
behavior which governs the longitudinal electrical conductivity $\sigma$.
\cite{zhang2,zhang3,Sheng92} In practice, an experimental detection of a
many-body correction to $\sigma$ is straightforward, while a measurement of a
small correction to the Hall coefficient $R_H$ in a metallic system would be a
challenging task.

A granular metal refers to a granular conductor with the dimensionless
intergrain tunneling conductivity $g_T$=$G_T$/($2e^2/h$)$\gg$1, where $G_T$ is
the intergrain tunneling conductance, $e$ is the electronic charge, and $h$ is
the Planck constant. Efetov, Beloborodov, and coworkers have lately carried
out a series of theoretical investigations in this regime. They found that the
Coulomb electron-electron ($e$-$e$) interaction effect governs the carrier
transport characteristics in the presence of granularity. Kharitonov and
Efetov predicted that, in the wide temperature interval of
$g_T\delta$$\alt$$k_BT$$\alt$$E_0$, $R_H$ should obey \cite{zhang4}
\begin{equation}\label{Eq.(Hall)}
R_H = \frac{1}{n^\ast e} \left[ 1 + \frac{c_d}{4\pi g_T} \ln \left(
\frac{E_0}{k_BT} \right) \right] \,,
\end{equation}
where $n^{\ast}$ is the effective carrier concentration, $c_d$ is a numerical
lattice factor, $\delta$ is the mean energy level spacing in the grain, $E_c$
is the charging energy, $E_0$=$\min(g_TE_c, E_\text{Th})$, and $E_\text{Th}$
is the Thouless energy.

In addition, Efetov and Tschersich \cite{zhang2} and Beloborodov {\it et al.}
\cite{zhang3} predicted that the intergrain $e$-$e$ interaction effect would
cause a longitudinal  electrical conductivity
\begin{equation}\label{Eq.(conductivity)}
\sigma = \sigma_0 \left[ 1 - \frac{1}{2\pi g_Td} \ln \left(
\frac{g_TE_c}{k_BT} \right) \right]
\end{equation}
in the temperature interval $g_T\delta$$<$$k_BT$$<$$E_c$, where
$\sigma_0$=$G_Ta^{2-d}$ is the tunneling conductivity between neighboring
grains in the absence of Coulomb interaction, $a$ is the radius of the grain,
and $d$ is the dimensionality of the granular array. We note that the
theories \cite{zhang2,zhang3,zhang4} have treated the simplest case of a
regular array consisting of equally sized spherical grains (while distributions in grain size, shape, and intergrain distance always exist in real systems).

Thus far, the theoretical prediction of Eq.~(\ref{Eq.(Hall)}) has not been
experimentally tested. The main reason is that the $n^\ast$ value is usually
very high ($\sim$$10^{28}$--$10^{29}$ m$^{-3}$) in those granular conductors
made of normal-metal grains, which leads to minute $R_H$ magnitudes.
Furthermore, the logarithmic correction term in Eq.~(\ref{Eq.(Hall)}) is
predicted to be $\lesssim$10\% of the total $R_H$ magnitude. Thus, an
experimental test of this theoretical $R_H$\,$\propto$\,ln$T$ law is nontrivial.

It is recently established that the indium tin oxide (ITO) material possesses
free-carrier-like electronic properties.
\cite{zhang11,zhang12,zhang13,zhang14} Their resistivities could be made as
low as $\rho$(300\,K)$\sim$100--200 $\mu \Omega$ cm.
\cite{zhang14,Hsu-prb10,Guo11} Therefore, one may consider growing ultrathin
($\sim$10 nm) ITO films to form granular arrays while achieving the
prerequisite condition $g_T$$\gg$1. Furthermore, since the $n^\ast$ magnitudes
in metallic ITO materials are $\sim$2--3 orders of magnitude lower than those
in typical metals, \cite{zhang14} one could expect relatively large values of
$R_H$. That is, the theoretical predication of Eq.~(\ref{Eq.(Hall)}), together
with that of Eq.~(\ref{Eq.(conductivity)}), may be tested by using granular
ITO films. In this work, we report the first experimental observation
of the $R_H$\,$\propto$\,ln$T$ law, as well as the $\sigma$\,$\propto$\,ln$T$
law, in a series of ultrathin ITO films which lie deep in the metallic regime.

\begin{table*}
\caption{\label{TableLi} Sample parameters for six ultrathin ITO
films. $T_s$ is the substrate temperature during deposition, $t$ is the mean film
thickness, $a$ is the mean grain radius determined from Fig.~\ref{FIGSEM}, and $n^\ast$ is the measured effective
carrier concentration. $c_d$ and $E_0$ ($\sigma_0$ and $g_T$) are adjusting parameters in Eq.~(\ref{Eq.(Hall)}) [Eq.~(\ref{Eq.(conductivity)})].
$\delta$ is the calculated mean energy level spacing. The theoretical Thouless energy
$E^\text{th}_{\text{Th}}$=$\hbar D/a^2$,  the experimental
charging energy $E_c$=10$k_BT^\ast$, and the theoretical charging energy
$E^\text{th}_c$=$e^2/(8\pi \epsilon_0 a)$. The standard deviations of $a$ for films Nos. 1--4 (5 and 6) are $\approx$20\% ($\approx$25\%). The uncertainties are $\approx$15\% in $E_0$ and $T^\ast$, a factor of $\sim$2 in $\delta$, $E^\text{th}_\text{Th}$ and $E^\text{th}_c$, and $\lesssim$5\% in other parameters.}

\begin{ruledtabular}
\begin{center}

\begin{tabular}{cccccccccccccccc}
Film & $T_s$ & $t$ & $a$ & $\rho$(300\,K) & $n^\ast$ & $T_\text{max}$ & $c_d$ & $E_0$ & $\delta$
& $E^\text{th}_\text{Th}$ & $T^\ast$ & $\sigma_0$ & $g_T$ & $E_c$ & $E_c^\text{th}$ \\
 & (K) & (nm) & (nm) & ($\mu \Omega$\,cm) & ($10^{27}$\,m$^{-3}$) & (K) & & ($10^{-21}$\,J) & ($10^{-24}$\,J)
 & ($10^{-22}$\,J) & (K) & ($10^5$\,S/m) & & ($10^{-21}$\,J) & ($10^{-21}$\,J) \\  \hline

1 & 610 & 9.7 &24 & 333 &1.1 &85 &0.72 & 1.3 & 3.8 & 1.5 & 46 & 3.2 & 13 &6.3 & 4.8\\
2 & 630 & 9.2 &28 & 302 &1.0 &73 &0.72 & 1.3 & 2.6 & 1.3 & 48 & 3.6 & 13 &6.6 & 4.1\\
3 & 650 & 11.3 &34 & 259 &1.1 &100 &1.1 & 2.0 & 1.7 & 1.0 & 49 & 4.2 & 23 &6.8 & 3.4\\
4 & 670 & 13.4 &38 & 226 &1.2 &120 &1.0 & 2.1 & 1.2 & 0.90 & 50 & 4.8 & 31 &6.9 & 3.0\\
5 & 650 & 7.6 &24 & 501 &0.73 &50 &0.70 & 0.99 & 5.1 & 1.2 & 55 & 2.4 & 7.4 &7.6 & 4.8\\
6 & 650 & 5.4 &22 & 839 &0.58 &25 &1.0 & 0.66 & 9.4 & 1.4 & 62 & 1.5 & 4.5 &8.6 & 5.2\\
\end{tabular}
\end{center}
\end{ruledtabular}
\end{table*}

Our ultrathin ITO films were deposited on glass substrates by the standard rf
sputtering method. A commercial Sn-doped In$_2$O$_3$ target (99.99\% purity,
the atomic ratio of Sn to In being 1:9) was used as the sputtering source. The
base pressure of the vacuum chamber was $\lesssim$8$\times$$10^{-5}$ Pa and
the sputtering deposition was carried out in an argon atmosphere (99.999\%) of
0.6 Pa. During the depositing process, the mean film thickness $t$, together with the substrate temperature $T_s$, was varied
to ``tune" the grain size $a$ and the intergrain
conductivity $g_T$ in each film. Hall-bar shaped samples (1.5 mm
wide and 1 cm long) were defined by using mechanical masks [see a schematic in the inset of Fig.~\ref{Li-Fig3}(a)]. The
thicknesses of the films were measured by the low-angle x-ray diffraction
(X'pertPRO multi purpose diffractometer). The surface morphologies of the
films were characterized by the scanning electron microscopy (SEM, Hitachi
S-4800). The four-probe electrical conductivity and Hall effect measurements
were carried out on a physical property measurement system (PPMS-6000, Quantum
Design). For $R_H$ measurements, in order to cancel out any
undesirable misalignment voltages and the thermomagnetic effect, a square-wave
current operating at a frequency of 8.33 Hz was applied and the magnetic field
was regulated to sweep from $-$2 to 2 T in a step of 0.05 T.

\begin{figure}[htp]
\begin{center}
\includegraphics[scale=0.55]{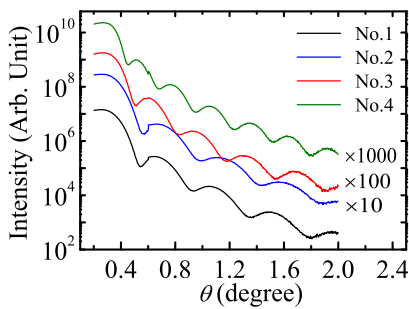}
\caption{(color online) Low-angle x-ray diffraction patterns for 4 ultrathin ITO films, as
indicated. The data for the films Nos. 2, 3, and 4 are offset for clarity.}
\label{FIGXRD}
\end{center}
\end{figure}

Figure~\ref{FIGXRD} shows the low-angle x-ray diffraction patterns of four
representative films, as indicated. The relation between the position of the
low-angle diffraction peak $\theta_m$ and film thickness $t$ is given by the
modified Bragg equation: \cite{zhang16}
$\sin^2(\theta_m)$=$(q\lambda /2t)^2$+$2\xi$,
where $\lambda$ is the x-ray wavelength (the Cu K$_\alpha$ radiation), $q$ is the integer reflection order number, and $\xi$ is the average
deviation of the refractive index from unity. The $t$ value of each film has
been determined from the linear regression of $\sin^2 (\theta_m)$ versus $q^2$
and is listed in Table~\ref{TableLi}.

\begin{figure}[htp]
\begin{center}
\includegraphics[scale=0.86]{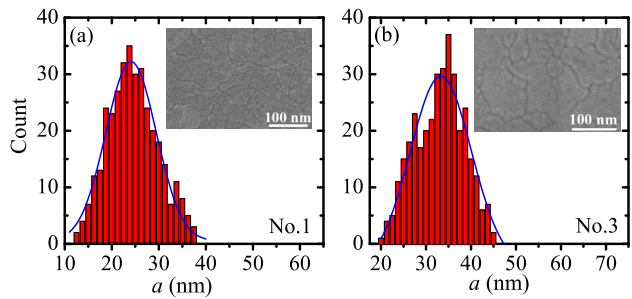}
\caption{(color online) Grain size distribution histograms for the ITO films (a) No. 1, and (b) No. 3. The insets show the corresponding SEM images.} \label{FIGSEM}
\end{center}
\end{figure}

Figures~\ref{FIGSEM}(a) and \ref{FIGSEM}(b) show the grain size distribution histograms for two representative films, as indicated. The insets show the corresponding SEM images which indicate irregular shape of individual grains. For each selected grain, we measured its size at 6 different locations and took the average value as the diameter of the grain, i.e., we modeled the grain as a disk-shaped grain with a height $t$. The distribution data were fitted with the standard Gaussian function (solid curves) to determined the mean grain radius $a$ in each film (see Table~\ref{TableLi}).

Figures~\ref{Li-Fig3}(a)--\ref{Li-Fig3}(d) plot the variation in Hall
coefficient $R_H$ with the logarithm of temperature for four representative
films, as indicated. All samples reveal negative $R_H$ at all $T$, indicating
electron conduction in the ITO material. \cite{zhang11,zhang14} It is clearly
seen that the Hall coefficient obeys the $R_H$\,$\propto$\,ln$T$ law between 2
K and $T_{\text{max}}$, where $T_{\text{max}}$ is the temperature below which
the logarithmic law holds. Our experimental $T_{\text{max}}$ values vary from
$\sim$50 to $\sim$120 K for films Nos. 1--5 (see Table~\ref{TableLi}).

Our measured $R_H$ variations with ln$T$ were least-squares fitted to the
predictions of Eq.~(\ref{Eq.(Hall)}) and the fitted results (straight solid
lines) are plotted in Figs. 3(a)--3(d). Note that the $n^\ast$ value in
Eq.~(\ref{Eq.(Hall)}) is given by the measured value between 180 and 250 K for
each film, and thus is not adjustable. (In this high $T$ regime, the
Coulomb interaction effect causes a negligible correction to $R_H$.
\cite{zhang4}) Our extracted $n^\ast$ values listed in Table~\ref{TableLi}
are in good accord with those previously measured in homogeneous ITO films.
\cite{zhang14} Among the three adjusting parameters $E_0$, $c_d$ and $g_T$,
the $g_T$ value can be independently determined by comparing the measured
$\sigma (T)$ with the prediction of Eq.~({\ref{Eq.(conductivity)}) (see below). Our fitted values of $E_0$ and $c_d$, together with the values of $g_T$ and $T_{\text{max}}$, are listed in Table~\ref{TableLi}.
Figures~\ref{Li-Fig3}(a)--\ref{Li-Fig3}(d) demonstrate that the
predictions of Eq.~(\ref{Eq.(Hall)}) can well described the experimental data
over $\sim$2 decades of temperature in all films, strongly suggesting that the
$e$-$e$ interaction effect does play an important role in the Hall transport
of granular metals. Our experimental results illustrate that the film thickness $t$, rather than the substrate temperature $T_s$,  plays a more dominant role in governing the variation in samples parameters.

\begin{figure}
\begin{center}
\includegraphics[scale=0.96]{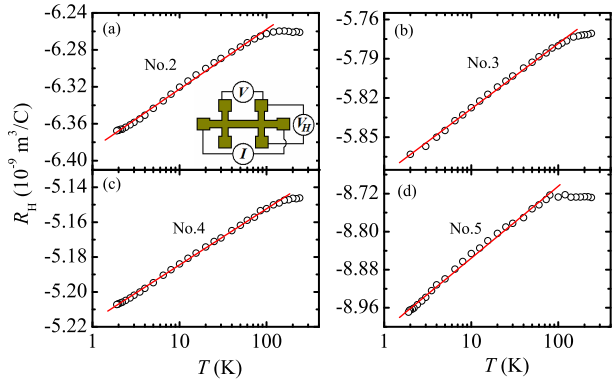}
\caption{(color online) Hall coefficient as a function of logarithm of
temperature for 4 ITO films, as indicated. The straight solid lines
are least-squares fits to Eq.~(\ref{Eq.(Hall)}). The inset in (a) depicts a schematic for our Hall-bar shaped sample.} \label{Li-Fig3}
\end{center}
\end{figure}


According to Kharitonov and Efetov, \cite{zhang4} Eq.~(\ref{Eq.(Hall)}) is
valid in the temperature range $g_T\delta$$\alt$$k_BT$$\alt E_0$. ($E_0$=$E_{Th}$ in this work.) The mean
energy level spacing $\delta$ in a single grain is given by $\delta$=$(\nu
V)^{-1}$, where $V$ is the volume of the grain, and $\nu$ is the electronic
density of states at the Fermi energy. Since ITO possesses a
free-electron-like bandstructure, \cite{zhang11,zhang14} we write
$\nu$=$m^\ast k_F/(\pi\hbar)^2$, where the Fermi wavenumber
$k_F$=$(3\pi^2n^\ast)^{1/3}$, and the effective electron mass
$m^\ast$=0.55\,$m_e$ ($m_e$ is the free-electron mass). \cite{zhang17} Our
calculated values of $\delta$ are listed in
Table~\ref{TableLi}. From Table~\ref{TableLi}, one readily obtains that the lower
limit in $T$ for Eq.~(\ref{Eq.(Hall)}) to be applicable is $g_T\delta
/k_B$$\sim$2--3 K in all samples. This is in good consistency with our
experimental observation. Furthermore, our extracted $T_{\text{max}}$ values
satisfy the condition $k_BT_{\text{max}}$$\lesssim$$E_0$. However, our
experimental $E_0$ values are $\sim$10 times greater than the theoretical
values of the Thouless energy $E^\text{th}_{\text{Th}}$=$\hbar D/a^2$,
\cite{zhang18} where $D$=$\sigma/(\nu e^2)$ is the electron diffusion
constant. This underestimate of $E^\text{th}_{\text{Th}}$ can be (partly) explained. To accurately evaluate $E^\text{th}_{\text{Th}}$ from $D$, one
should have used the intrinsic conductivity $\sigma_\text{grain}$ of an
individual ITO grain, instead of using the measured $\sigma$ of the film.
Therefore, the $E^\text{th}_{\text{Th}}$ values listed in Table~\ref{TableLi}
only represent the lower bounds, because $\sigma_\text{grain}$$>$$\sigma$ in a
granular array. The fact that our grains are disk-shaped but not spherical
could have introduced additional uncertainties in the estimate. In short, our
measured ln$T$ behavior of $R_H$ can be satisfactorily described by
Eq.~(\ref{Eq.(Hall)}).

\begin{figure}
\begin{center}
\includegraphics[scale=0.96]{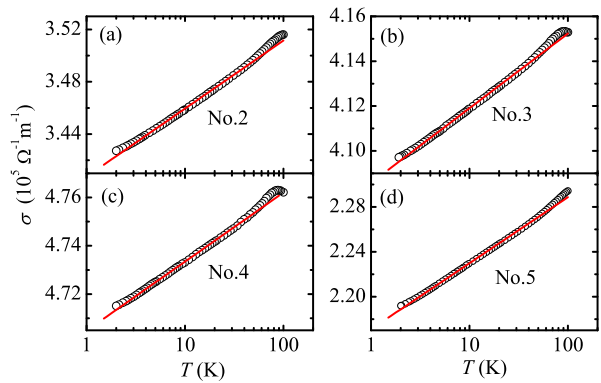}
\caption{(color online) Variation in longitudinal electrical conductivity with logarithm of
temperature for 4 ITO films measured in a
perpendicular magnetic field of 7 T. The straight solid lines are
least-squares fits to Eq.~(\ref{Eq.(conductivity)}). } \label{Li-Fig4}
\end{center}
\end{figure}

As mentioned, if the Coulomb interaction effect dominates the electron
dynamics in granular ITO films, our measured $\sigma (T)$ should follow the
predications of Eq.~(\ref{Eq.(conductivity)}) in the temperature interval
$g_T\delta$$<$$k_BT$$<$$E_c$. According to Efetov and Tschersich \cite{zhang2}
and Beloborodov {\it et al.}, \cite{zhang3,Beloborodov-prb04} the
weak-localization (WL) effect originally formulated for \textit{homogeneous} systems
\cite{Bergmann,zhang19} should be suppressed at $T$$>$$g_T \delta$/$k_B$.
Empirically, it has been found that the WL effect in thick ITO
films could persist up to several tens K. \cite{zhang17} In order to fully
exclude any residual WL effect on $\sigma$, we have measured
$\sigma (T)$ of our ultrathin ITO films in a perpendicular magnetic field $B$ of 7
T. \cite{Beloborodov-prb04} Our results for 4 representative films
are plotted in Fig.~\ref{Li-Fig4}. Our measured $\sigma$ data are compared
with Eq.~(\ref{Eq.(conductivity)}) and the least-squares fitted results are
plotted as the straight solid lines. Note that the prediction of
Eq.~(\ref{Eq.(conductivity)}) is valid in any $B$ as long as
$\omega_c \tau$$<$1, where $\omega_c$ is the cyclotron frequency and $\tau$ is
the electron mean free time. \cite{zhang3} In our fitting processes,
$\sigma_0$ and $g_T$ are treated as adjusting parameters, and the charging
energy is taken to be $E_c$$\approx$10$k_BT^\ast$, \cite{zhang10} where
$T^\ast$ is the temperature below which the $\sigma$\,$\propto$\,ln$T$ law
holds (see Table~\ref{TableLi}). \cite{T*} The array dimensionality $d$=2 in this work, since our ultrathin films
are nominally covered with only one layer of ITO grains. Our fitted values of
$\sigma_0$ and $g_T$ are listed in Table~\ref{TableLi}. Figure~\ref{Li-Fig4}
indicates that our experimental data between $\sim$3 K and $T^\ast$ are well
described by Eq.~(\ref{Eq.(conductivity)}). The values of
$E_c$$\approx$10$k_BT^\ast$ are comparable to the theoretical estimates
$E^\text{th}_c$=$e^2/(8\pi \epsilon_0 a)$ within experimental uncertainties,
where $\epsilon_0$ is the permittivity of vacuum. For films Nos. 1--4, our
extracted $g_T$ values are far greater than 1, while for the film No. 6,
$g_T$$\simeq$4.5. This latter value suggests that even the thinnest film No. 6 lies in
the metallic region. Thus, Eq.~(\ref{Eq.(Hall)}) and
Eq.~(\ref{Eq.(conductivity)}) are safely applicable for our films.

\begin{figure}
\begin{center}
\includegraphics[scale=0.78]{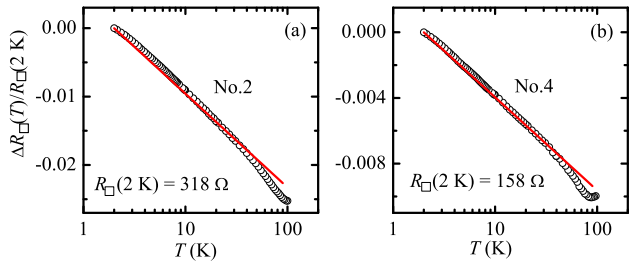}
\caption{(color online) Normalized sheet resistance, $\Delta R_\square
(T)/R_\square (2\,\text{K})$, as a function of logarithm of temperature in a
perpendicular magnetic field of 7 T for two ITO films, as indicated. The straight
solid lines are least-squares fits to the two-dimensional homogeneous $e$-$e$ interaction theory (see text).}
\label{Li-Fig5}
\end{center}
\end{figure}

It is well known that the $e$-$e$ interaction effect also results in a small
ln$T$ correction to longitudinal conductivity (or resistivity) in
two-dimensional systems at low $T$. \cite{zhang19} The correction to
the sheet resistance, $R_\square$, due to the $e$-$e$ interaction effect in a
homogeneous, weakly disordered film is given by \cite{zhang19,zhang22} $\Delta R_\square (T)/R_\square (T_0)$= $-$$(e^2/2\pi^2 \hbar)$(1$-$$3\widetilde{F}$/4)$R_\square (T_0)$\,ln$(T/T_0)$,
where $\widetilde{F}$ is a screening factor, and $T_0$ is an
arbitrary reference temperature. Figure~\ref{Li-Fig5} shows the normalized
sheet resistance, $\Delta R_\square (T)/R_\square (2\,\text{K})$= $[R_\square
(T)$$-$$R_\square(2\,\text{K})$]/$R_\square (2\,\text{K})$, for the films Nos.
2 and 4 measured in a perpendicular $B$ of 7 T as a function of
ln$T$. (The rest films behave in a similar manner.) The straight solid lines
are the least-squares fits to this theory. Although an
approximate ln$T$ regime seems to exist for $T$ below $\sim$35 K, our fitted
values of $\widetilde{F}$ are $-$0.61 and $-$0.35 for the films Nos. 2 and 4,
respectively. Since this $e$-$e$ interaction theory requires
that 0$\lesssim$$\widetilde{F}$$\lesssim$1, \cite{zhang19} the seemingly good fits shown in
Fig.~\ref{Li-Fig5} are thus spurious. That is, our measured
$\sigma$\,$\propto$\,ln$T$ law in ultrathin ITO films cannot be ascribed to
the conventional $e$-$e$ interaction effect in homogeneous systems.

In conclusion, we have studied the temperature dependences of Hall coefficient
and longitudinal conductivity in a series of ultrathin indium tin oxide films.
The films were specifically made granular, while possessing overall metallic
behavior. We observed the robust $R_H$\,$\propto$\,ln$T$
law, together with the $\sigma$\,$\propto$\,ln$T$ law, over nearly 2 decades
of temperature below $\sim$100 K. Our results are fairly quantitatively
understood within these recent theoretical frameworks of the electron-electron interaction effect in the presence of granularity. It is meditative that these theories which are formulated based on a regular array of spheres can be so successfully applied to explain real systems where distributions in grain size, shape, and intergrain distance exist.

This work was supported by the Key Project of Chinese MOE through Grant No. 109042 and Tianjin City NSF through Grant No.
10JCYBJC02400 (Z.Q.L.), and by the Taiwan NSC through Grant No. NSC 99-2120-M-009-001 and the MOE ATU Program (J.J.L.).


\begin{thebibliography}{00}
\bibitem{T*} Our $T^\ast$ values are lower than the corresponding $T_\text{max}$ values in those most metallic films Nos. 1--4, because that the contribution from the electron-phonon scattering to $\sigma$ (but not to $R_H$) becomes progressively notable as $T$$\gtrsim$50 K.
\end{thebibliography}
\end{document}